*Chapter 1*

# IIGHGINT: A GENERALIZATION TO THE MODIFIED GHG INTENSITY UNIVERSAL INDICATOR TOWARD A PRODUCTION/CONSUMPTION INSENSITIVE BORDER CARBON TAX

***Reza Farrahi Moghaddam,*** [*] ***Fereydoun Farrahi Moghaddam, and Mohamed Cheriet***
Synchromedia Lab, École de technologie supérieure (ETS),
University of Quebec (UduQ), Montreal, QC, Canada

**Abstract**

A global agreement on how to reduce and cap human footprint, especially their GHG emissions, is very unlikely in near future. At the same time, bilateral agreements would be inefficient because of their neural and balanced nature. Therefore, unilateral actions would have attracted attention as a practical option. However, any unilateral action would most likely fail if it is not fair and also if it is not consistent with the world trade organization's (WTO's) rules, considering highly heterogeneity of the global economy. The modified GHG intensity (MGHGINT) indicator, hereafter called Inequality-adjusted Production-based GHGINT (IPGHGINT), was put forward to address this need in the form of a universal indicator applicable to every region regardless of its economic and social status. Nonetheless, the original MGHGINT indicator ignores hidden consumption-related emissions, and therefore it could be unfair to some production-oriented regions in the current bipolar production/consumption world. Here, we propose two generalizations, called Inequality-adjusted Consumption-based GHGINT (ICGHGINT) and Inequality-adjusted Production/Consumption-Insensitive GHGINT (IIGHGINT), to the IPGHGINT in order to combine both production and consumption emissions in a unified and balanced manner. The impact of this generalizations on the associated border carbon tax rates is evaluated in order to validate their practicality.

[*] Corresponding author's e-mail address: imriss@ieee.org (Reza Farrahi Moghaddam)

# 1. Introduction

Global agreement on ubiquitous actions toward a worldwide sustainable future has been urged by almost all parties and actors across the world taking into account the considerable time that has been lost in the past decades of solo actions, shifts, and leakages. The source of this situation could be traced back to the general doubt about the possibility of an indicator that can assess and evaluate actions of all countries and regions across the world regardless of their fundamental differences in terms of development, life style, and population, among other parameters. In contrast, we think that such a universal indicator is possible, and among various strategies to address this challenge, development of an universal indicator of environmental impacts was proposed in [3] by introducing a modified version of the Greenhouse Gases (GHG) emissions intensity (GHGINT) indicator that combines the GDP, population, and inequality-adjusted development of countries in order to make all countries comparable in terms of their GHG emissions footprint. In addition, considering a green and a red scenarios for the global level of GHG emissions, that proposed indicator was leveraged into carbon border tax (BCT) or carbon border adjustment (BCA) rates calculated for and applicable to individual countries [3]. The key factor of such BCT/BCA mechanisms is in their low-cost administration because they are applied to all products of a country regardless of their individual embodied emissions. The administrative costs, and also the uncertainty and the delay associated with calculating the embodied emissions of a product in other mechanisms can be seen as one of their main challenges toward their implementation. The BCT/BCA rates could be used by any country that has enforced an internal equivalent emission control mechanism, and therefore can be the building block of any unilateral action against environmental footprint of other countries while conforming to the world trade organization's (WTO's) rules. Although the details of exact conformation to the WTO's rules require other studies, the possibility of such unilateral actions based on the proposed universal indicator of GHG emissions footprint shows the the great potential and importance of the universal indicators in bringing novel and fair options to the negotiation tables toward a global agreement.

Although the proposed modification to the GHG emissions intensity in [3] showed that big global polluters, such as the USA and China, can be assessed fairly on the same scale, and then can be penalized using fair BCT rates, it is silent with respect to those emissions that are occurred in a region to serve a customer in another region. These imported emissions seem to be considerable, and it has been observed that up to %24 of china's $CO_2$ emissions were embodied in its exports in 2001 [9]. This can be the same for many other countries. In other words, the business attraction that consumers of a region create can be the main motive for the emissions in another region. Although penalizing the emitter seems a fair practice, penalizing the consumer region is also seems necessary because the economic drive they produce can cancel out the penalty imposed on the emitter and therefore make their business profitable. At the same time, other producers in other regions (including those in the consumer's region) may face an unfair situation where tight regulations in their region make it unprofitable for them to participate in competition to provide the product or service to the consumer. This shows the necessity to modify the proposed modified GHGINT in [3], called hereafter Inequality-adjusted Production-based GHGINT (IPGHGINT), to a higher level in order to include the consumption-related emissions.

In a bigger picture, we consider a road map toward more fair universal indicators of GHG emissions, as shown in Figure 1. The starting point of this road map is the original GHGINT indicator that considers emissions per unit of production (measured as GDP). The next step was to consider the GDP (PPP)[1] instead the GDP in the definition of the GHGINT, in order to arrive to the PGHGINT indicator that corrects the indicator values with respect to the currency exchange rates bias. The next step was taken in [3] to include population activities and also unequal development across a country, shown in the figure as the IPGHGINT indicator. As discussed before, the next obvious step would be to include the consumption role and impact in the universal GHG emissions indicator, which is the focus of this paper and is highlighted as the ICGHGINT indicator in Figure 1. Beyond this work, we see a potential step toward more fair and universal indicator, among other potential modifications, to be the inclusion of geographical and climate characteristics of countries and regions into the indicator. We think that these factors are contributing a considerable bias in the indicators, and therefore could have led to masking of under development in some regions that is in contrast with the ultimate goal of having ubiquitous development across the globe. This will be addressed in future studies.

It is worth noting that despite using the BCT/BCA rates in this work and also in [3], the proposed universal indicators can work and can be used completely independent from any tax system. In other words, these indicators can be used to develop any mechanism, including emission trading and exchange systems. However, as a use case, and also because of unique benefits of the BCT/BCA mechanisms mentioned before, we use these enforcement mechanisms in this study to show the potential of the proposed ICGHGINT universal indicator.

The paper is organized as follows. The data used in this study and their sources are discussed in section 2. The definition of the Inequality-adjusted Production-based GHGINT (IPGHGINT) indicator is provided in section 3. The generalization to Inequality-adjusted Consumption-based GHGINT (ICGHGINT) is provided in section 4. The impact of considering a production/consumption-insensitive indicator, the IIGHGINT, and comparison with other indicators including the GHGINT, the GHG emissions per capita (GHGpCapita), the IPGHGINT, and the ICGHGINT are discussed and presented in section 5. Finally, the conclusions are provided in section 6.

[1]GDP at purchasing power parity (PPP) exchange rates.

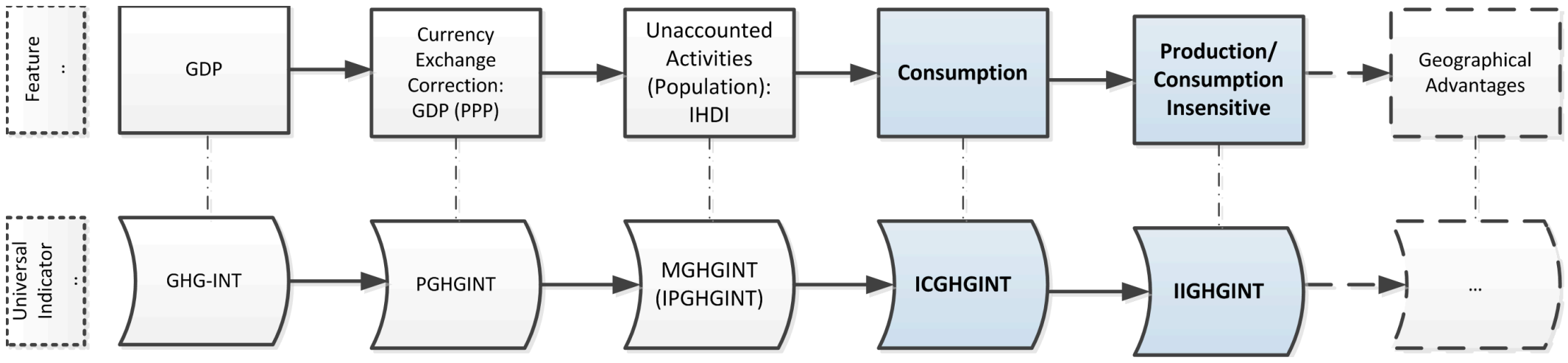

Figure 1. The proposed road map toward more fair and universal emission intensity indicators. This paper covers the highlighted steps of introducing the ICGHGINT and the IIGHGINT indicators by considering the consumption-based emissions and insensitivity with respect to production and consumption emissions.

## 2. Data sources

In this work, we use the $CO_2$ and non-$CO_2$ emissions data of 2009. For the purpose of consistency, the production and consumption $CO_2$ emissions of countries are obtained from the Organization for Economic Cooperation and Development (OECD) database on the carbon dioxide emissions embodied in international trade, the STAN IO GHG database.[2] This database considered 57 OECD and non-OECD countries. The consumption-based $CO_2$ emissions of a country is defined as:

$$\mathrm{EM}_{i,y}^{CO_2,\mathrm{CONS}} = \mathrm{EM}_{i,y}^{CO_2,\mathrm{CATEG}} + \mathrm{EM}_{i,y}^{CO_2,\mathrm{DIRECT}} + \mathrm{EM}_{i,y}^{CO_2,\mathrm{ROAD}} + \mathrm{EM}_{i,y}^{CO_2,\mathrm{FOT}} \tag{1}$$

where $\mathrm{EM}_{i,y}^{CO_2,\mathrm{CATEG}}$, $\mathrm{EM}_{i,y}^{CO_2,\mathrm{DIRECT}}$, $\mathrm{EM}_{i,y}^{CO_2,\mathrm{ROAD}}$, and $\mathrm{EM}_{i,y}^{CO_2,\mathrm{FOT}}$ are the carbon dioxide emissions by household consumption type, the direct carbon dioxide emissions by households, the households' carbon dioxide emissions by road transport, and the carbon dioxide emissions by other expenditures of country $i$ in year $y$. As an example, for the USA, the $\mathrm{EM}_{i,y}^{CO_2,\mathrm{CATEG}}$ counted for %62.42 of its consumption-based $CO_2$ emissions, while the second rank was the $\mathrm{EM}_{i,y}^{CO_2,\mathrm{FOT}}$ with a share of %23.16.

In contrast, for the non-$CO_2$ emissions, which are mostly related to agricultural activities, we only consider the production emissions. This was adapted in order to avoid any additional advantage to agricultural-exporting regions because otherwise this could result in unprofitability of agricultural activities in the importing regions and therefore change in the landuse, which could in turn have a negative impact on the global food source portfolio and the global food security. The non-$CO_2$ emissions are obtained from the US Energy Information Administration and World Bank databases [14].[3] Then, the data related to the 57 countries of the STAN IO GHG database are extracted and considered in the calculations.

[2] http://stats.oecd.org/Index.aspx?DataSetCode=STAN_IO_GHG

[3] http://data.worldbank.org/indicator/EN.ATM.METH.KT.CE/countries?page=1, http://data.worldbank.org/indicator/EN.ATM.NOXE.KT.CE/countries, and http://data.worldbank.org/indicator/EN.ATM.GHGO.KT.CE/countries?page=1.

In summary, we use production-based $CO_2$ emissions (for the IPGHGINT indicator) and consumption-based $CO_2$ emissions (for the ICGHGINT indicator), and we will use only the production-based $CH_4$, $N_2O$, and HFS[4] emissions for the non-$CO_2$ emissions. The total GHG emissions of a region is the sum of $CO_2$ and non-$CO_2$ emissions in equivalent $CO_2$ weights.

The economic and social indicators, such as GDP at purchasing power parity exchange rates (PPP) [4, 13], population, and the Inequality-adjusted Human Development Index (IHDI) [1, 12], were obtained from the United Nations Statistics and Research Database (UNdata), the International Monetary Fund (IMF), and the United Nations Development Programme (UNDP) database covering the same period of time.[5] The Green and Red scenarios, which will be referred to in section 4., were taken from [3], and are basically built based on the B1 Asian-Pacific Integrated Model (AIM) and the A1B AIM scenarios of the Intergovernmental Panel on Climate Change (IPCC) [2, 5, 6, 8].[6]

It is worth mentioning that although we use the OECD database for production- and consumption-based $CO_2$ emissions, other approaches such as the Emissions Embodied in Bilateral Trade (EEBT) method [7, 10, 11], which focuses on the difference in the gross exported and imported emissions instead of consumption, can be used. The comparison of consumption-based and the EEBT-based approaches is beyond this work, and we will focus on the consumption-based emissions from here on.

The use of names of countries in the figures and tables serves only to identify world regions, and does not imply the expression of any opinion on the legal status of any country or its authorities, or concerning its boundaries. Also, a set of selected countries are considered as the countries of interest (COIs). The COIs is composed of countries with global or regional influence, and includes *Australia, Brazil, Canada, China, France, Germany, Greece, Hong Kong, India, Japan, Russia, South Africa, Switzerland, and the United States*.

# 3. Inequality-adjusted Production-based GHG Intensity (IPGHGINT) indicator

The first modified GHG intensity was proposed in [3]. In that work, in order to develop universal indicator that can be applied to all regions and countries across the world, the GHG intensity was modified toward local activities in each region that may be unaccounted in its GDP. This has been achieved by two modification. First, the GDP is replaced with the GDP (PPP), which stands for the GDP at purchasing power parity (PPP) exchange rates [4]. This allows to normalize currency exchange rate related discrepancies. To be precise, we use the PGHGINT indicator to represent this modification:

$$\mathrm{PGHGINT}_{i,y} = \frac{\mathrm{EM}_{i,y}}{\mathrm{GDP}^{\mathrm{PPP}}_{i,y}}, \tag{2}$$

where $\mathrm{PGHGINT}_{i,y}$ is the production-modified GHG intensity indicator of the country $i$ in year $y$ after considering its GDP (PPP), $\mathrm{GDP}^{\mathrm{PPP}}_{i,y}$, in the formulation, and $\mathrm{EM}_{i,y}$ is the total

[4] HPS stands for HFC, PFC and SF6 gases, collectively.

[5] http://hdrstats.undp.org/en/indicators/73206.html and http://www.imf.org/external/pubs/ft/weo/2011/01/weodata/index.aspx.

[6] http://sres.ciesin.org/final_data.html

GHG emissions emitted in the country $i$ in year $y$:

$$\mathrm{EM}_{i,y} = \mathrm{EM}_{i,y}^{CO_2} + \mathrm{EM}_{i,y}^{CH_4} + \mathrm{EM}_{i,y}^{N_2O} + \mathrm{EM}_{i,y}^{\mathrm{HPS}}, \tag{3}$$

where $\mathrm{EM}_{i,y}^{CO_2}$, $\mathrm{EM}_{i,y}^{CH_4}$, $\mathrm{EM}_{i,y}^{N_2O}$, and $\mathrm{EM}_{i,y}^{\mathrm{HPS}}$ stand for carbon dioxide ($CO_2$), methane ($CH_4$), nitrous oxide ($N_2O$), and HPS emissions occurred in the jurisdiction of the country $i$ in year $y$ (all measured in equivalent gigatonnes of $CO_2$ ($GtCO_2e$)). Please refer to section 2. for details on the sources of data used.

The second, and also the major modification step considered in [3] was adjusting the GHG intensity of a country based on the population's activities of that country. However, it was also argued that the population's activities of a region could be highly different from those of the other region in that sense that not all citizens could have had the same level of access to developed infrastructure. In other words, a citizen cannot perform an "activity" if their region has not developed to support that activity. Even if the development is in place, a citizen may not participate in an activity because of limited access to the associated support. To account for these factors that reduce the level of activity, the Inequality-adjusted Human Development Index (IHDI) [1] was used in [3] in order to modify the PGHGINT indicator in such a way that accounts for the inequality-adjusted internal activities of a region or country. The resulting new GHG intensity indicator was called the Modified GHG intensity in that work. However, we will use the Inequality-adjusted production-modified GHG intensity (IPGHGINT) indicator to represent this indicator here on:

$$\mathrm{IPGHGINT}_{i,y} = \frac{\mathrm{EM}_{i,y}}{\mathrm{IHDIGDP}_{i,y}^{\mathrm{PPP}}}, \tag{4}$$

where $\mathrm{IPGHGINT}_{i,y}$ is the Inequality-adjusted production-modified GHG intensity indicator of the country $i$ in year $y$, and $\mathrm{IHDIGDP}_{i,y}^{\mathrm{PPP}}$ is its IHDI-adjusted GDP (PPP) defined as [3]:

$$\mathrm{IHDIGDP}_{i,y}^{\mathrm{PPP}} = \left(\frac{Z}{2}\right) \frac{\mathrm{EM}_{i,y}}{\mathrm{GDP}_{i,y}^{\mathrm{BAL}} + \mathrm{IHDIxCapita}_{i,y}^{\mathrm{BAL}}}, \tag{5}$$

where $\mathrm{IHDIxCapita}_{i,y}$ is the multiplication of the IHDI of a country $i$ in year $y$ with its "snapshot" population in that year:[7]

$$\mathrm{IHDIxCapita}_{i,y} = \mathrm{IHDI}_{i,y}\mathrm{Capita}_{i,1990}. \tag{6}$$

Also, $Z$ is a factor of normalization in order to ensure the sum of $\mathrm{IHDIGDP}_{\cdot,1990}^{\mathrm{PPP}}$ of all countries is equal to the sum of $\mathrm{GDP}_{\cdot,1990}^{\mathrm{PPP}}$ of all countries, and $\mathrm{GDP}_{i,y}^{\mathrm{BAL}}$ and $\mathrm{IHDIxCapita}_{i,y}^{\mathrm{BAL}}$ are defined as follows:

$$\mathrm{GDP}_{i,y}^{\mathrm{BAL}} = \mathrm{GDP}_{y}^{\mathrm{PPP,\,Max}} \frac{\mathrm{GDP}_{i,y}^{\mathrm{PPP}}}{\mathrm{GDP}_{y}^{\mathrm{PPP,\,IHDI}}}, \tag{7}$$

$$\mathrm{IHDIxCapita}_{i,y}^{\mathrm{BAL}} = \mathrm{GDP}_{y}^{\mathrm{PPP,\,Max}} \frac{\mathrm{IHDIxCapita}_{i,y}}{\mathrm{IHDIxCapita}_{y}^{\mathrm{Max}}}, \tag{8}$$

[7]The snapshot population of a country in any year is defined as its population in year 1990. The year 1990 is selected as the reference year because it is a common reference year in many studies and indicators [3].

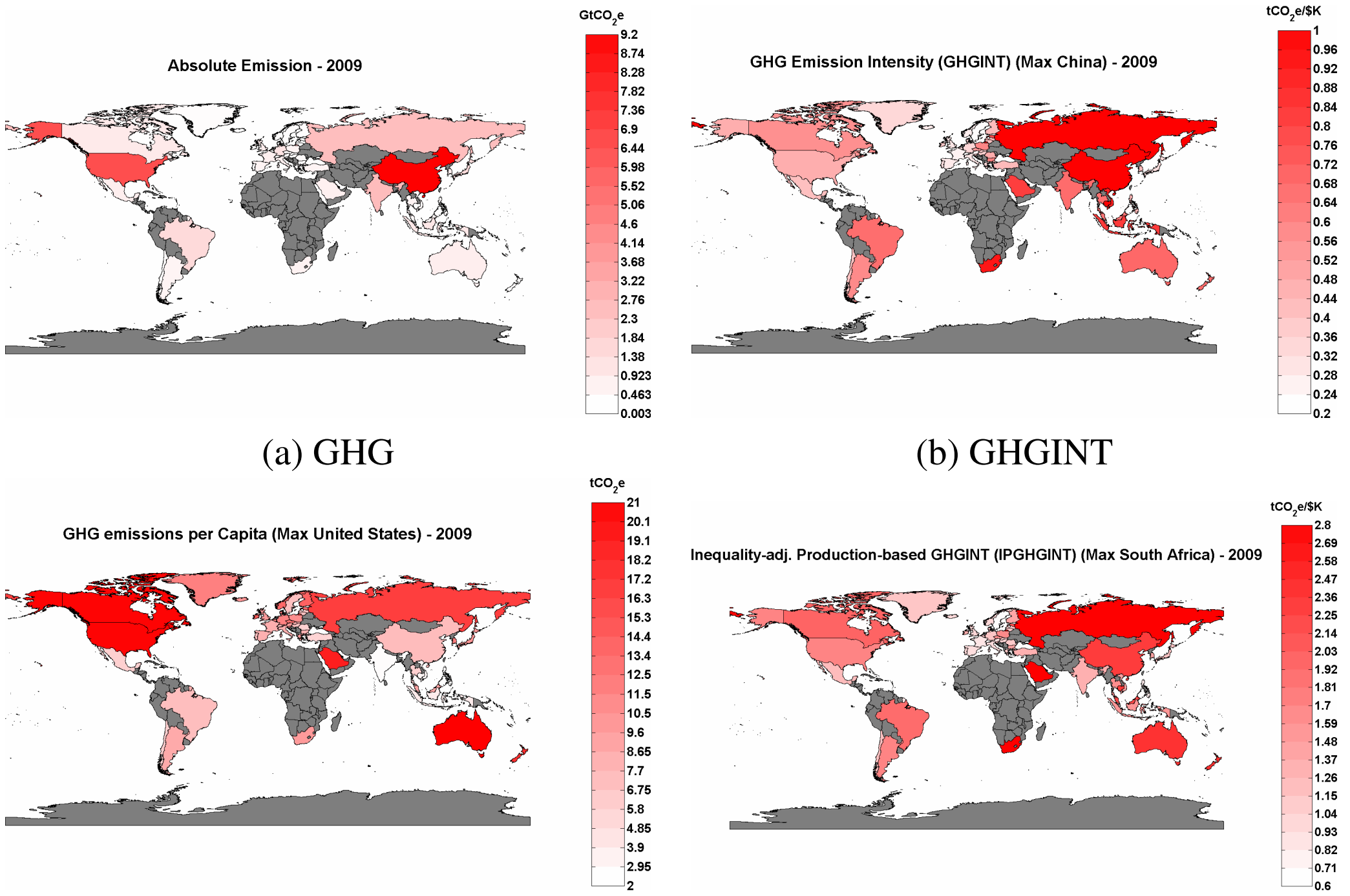

(a) GHG (b) GHGINT

(c) GHGpCapita (d) IPGHGINT

Figure 2. The visual comparison of four indicators when production-based $CO_2$ emissions are considered in the calculations.

where $GDP_y^{PPP,\,Max}$ is the maximum value of $GDP_{i,y}^{PPP}$ in year $y$, $GDP_y^{PPP,\,IHDI}$ is the $GDP^{PPP}$ of that country that its IHDIxCapita is maximum in year $y$: $IHDIxCapita_y^{Max}$.

Figure 3. provides a visual comparison of four indicators across the globe for the countries presented in the STAN IO GHG database in 2009. The indicators are the total GHG emissions, the GHGINT, the GHGpCapita, and the IPGHGINT receptively. Although the balance alternates between the two main emitters, especially when switching between the GHGINT and the GHGpCapita indicators, the EU region stays at low footprint side with respect to all indicators. Even in the case of the IPGHGINT indicator that provides a balance between the main emitters, the EU region keeps its low footprint stand. This brings a question about the necessity of a consumption-based indicator considering the amount of 823.1 $MtCO_2e$ difference in the EU's consumption and production emissions according to the STAN IO GHG database. This question is answered using the ICGHGINT indicator in the next section. It is worth nothing that in terms of production-based emissions, the COI countries with the worst scores with respect to the GHGINT, the GHGpCapita, and the IPGHGINT indicators are China, the USA, and South Africa, respectively.

# 4. Inequality-adjusted Consumption-based GHG Intensity (ICGHGINT) indicator

The Inequality-adjusted Consumption-based GHG-INT (ICGHGINT) indicator is defined as follows:

$$\mathrm{ICGHGINT}_{i,y} = \frac{\mathrm{EM}_{i,y}^{\mathrm{CONS}}}{\mathrm{IHDIGDP}_{i,y}^{\mathrm{PPP}}}, \tag{9}$$

where $\mathrm{EM}_{i,y}^{\mathrm{CONS}}$ is the consumption-based GHG emissions of the country $i$ in year $y$:

$$\mathrm{EM}_{i,y}^{\mathrm{CONS}} = \mathrm{EM}_{i,y}^{CO_2,\mathrm{CONS}} + \mathrm{EM}_{i,y}^{CH_4} + \mathrm{EM}_{i,y}^{N_2O} + \mathrm{EM}_{i,y}^{HPS}, \tag{10}$$

where $\mathrm{EM}_{i,y}^{CO_2,\mathrm{CONS}}$ is the consumption-based $CO_2$ emissions of the country $i$ in year $y$ obtained from the STAN IO GHG database. According to this data base there has been a 1516.2 $MtCO_2e$ additional emissions associated to the OECD countries, imported from non-OECD countries, when consumption is considered.[8] This is equal to %12.61 of their production-based $CO_2$ emissions. Despite this difference, there is no noticeable visual difference in the four indicators, as shown in Figure 4. And, as respected, the lead worst COI countries are China, the USA, and South Africa as shown in the figure. In order to show the impact of considering consumption-based emissions, the differences between the two case of production and consumption are shown in Figure 4. Interestingly, the EU regions appears as a worst region with respect to all indicators. In particular, Greece shows an increase of 0.11 $tCO_2e$/\$K with respect to the GHGINT, and Switzerland receives an increase of 3.3 $CO_2e$ with respect to the GHGpCapita. The first rank with respect to the $(\mathrm{ICGHGINT} - \mathrm{IPGHGINT})$ was Hong Kong with an increase of 0.39 $tCO_2e$/\$K.

The comparison of the GHGINT, the IPGHGINT, and the ICGHGINT indicators for the COIs is provided in Tables 1 and 2. The reason for having two GHGINT indicators, as provide in Table 1, is that we can use either the production-based or the consumption-based emissions as its numerator. As can be seen from the tables, the IPGHGINT and the ICGHGINT provide a more consistent intensity scores for the COIs regardless of their fundamental differences in terms production, development, and consumption. In particular, the ICGHGINT brings the European and other COIs closer to each other in terms of intensity by removing the bias related to consumption.

| COI | GHGINT | COI | GHGINT |
|---|---|---|---|
| **Australia** | **0.668** | **Hong Kong** | **0.162** |
| **Brazil** | **0.651** | **India** | **0.657** |
| **Canada** | **0.545** | **Japan** | **0.301** |
| **China** | **1.011** | **Russia** | **1.09** |
| **France** | **0.237** | **South Africa** | **0.929** |
| **Germany** | **0.328** | **Switzerland** | **0.167** |
| **Greece** | **0.32** | **United States** | **0.449** |

(a) Production-based

| COI | GHGINT | COI | GHGINT |
|---|---|---|---|
| **Australia** | **0.704** | **Hong Kong** | **0.262** |
| **Brazil** | **0.671** | **India** | **0.641** |
| **Canada** | **0.558** | **Japan** | **0.333** |
| **China** | **0.904** | **Russia** | **0.877** |
| **France** | **0.308** | **South Africa** | **0.846** |
| **Germany** | **0.37** | **Switzerland** | **0.248** |
| **Greece** | **0.428** | **United States** | **0.485** |

(b) Consumption-based

Table 1. The GHGINT of the COIs in the two cases of production-based and consumption-based $CO_2$ emissions.

[8]For which 823.1 $MtCO_2e$ is associated with the EU region.

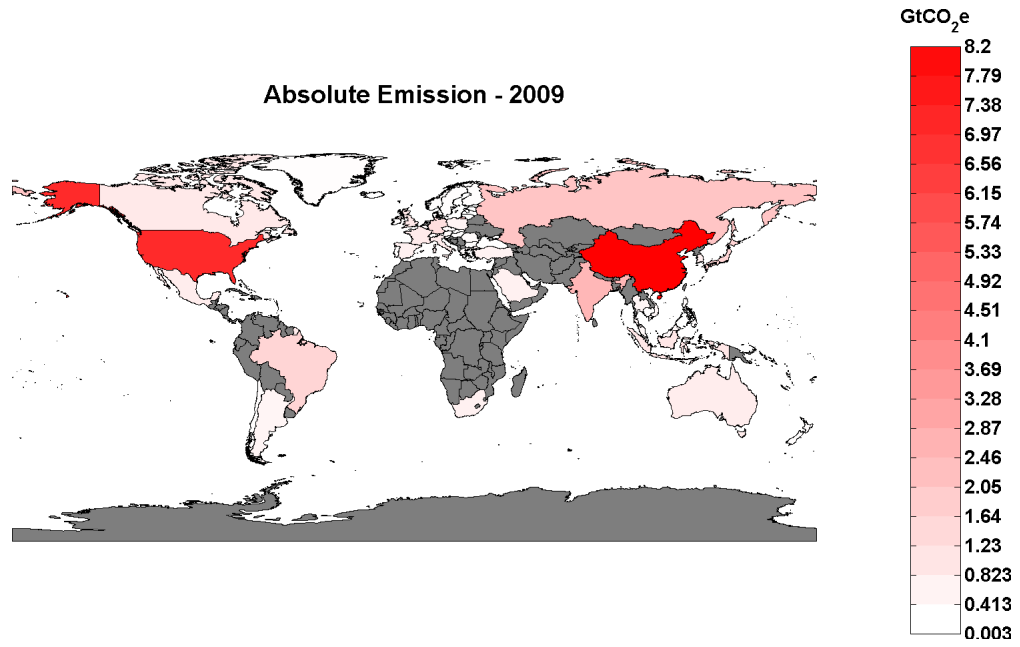

(a) GHG

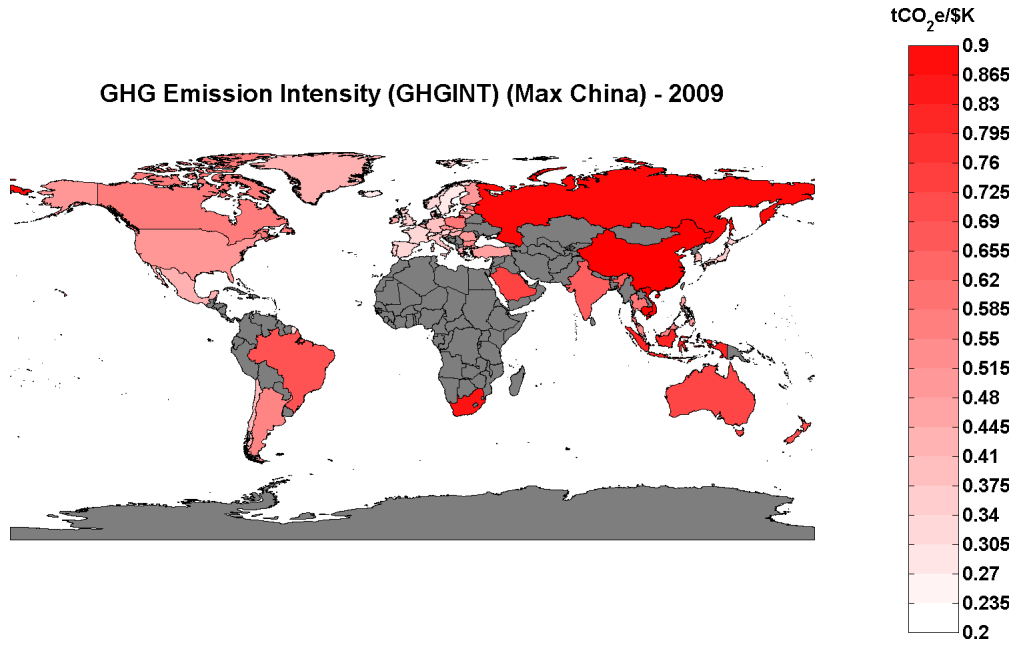

(b) GHGINT

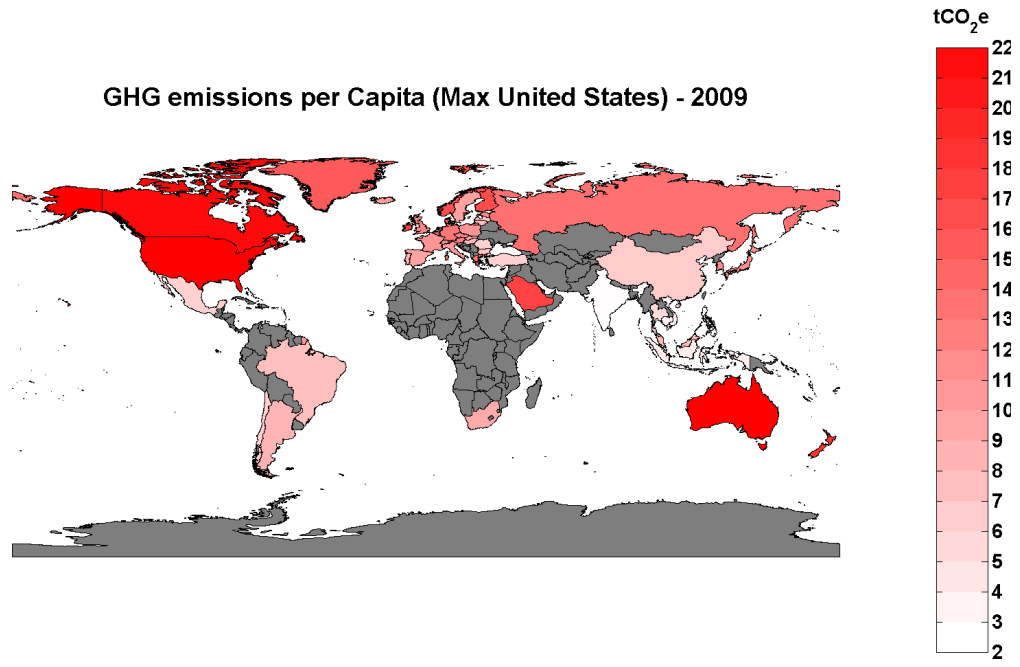

(c) GHGpCapita

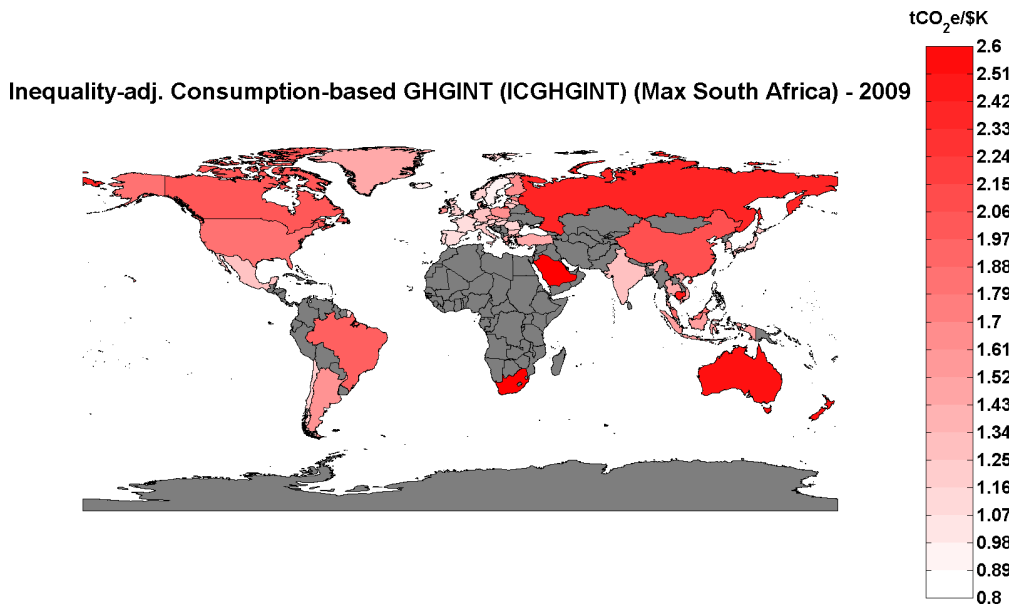

(d) ICGHGINT

Figure 3. The visual comparison of four indicators when consumption-based $CO_2$ emissions are considered in the calculations.

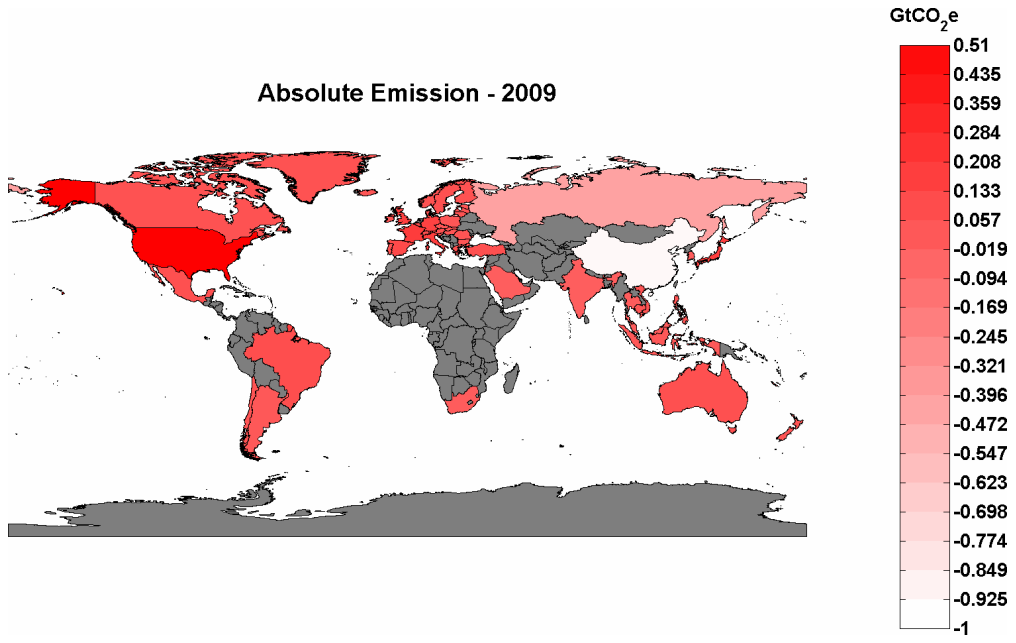

(a) Δ GHG

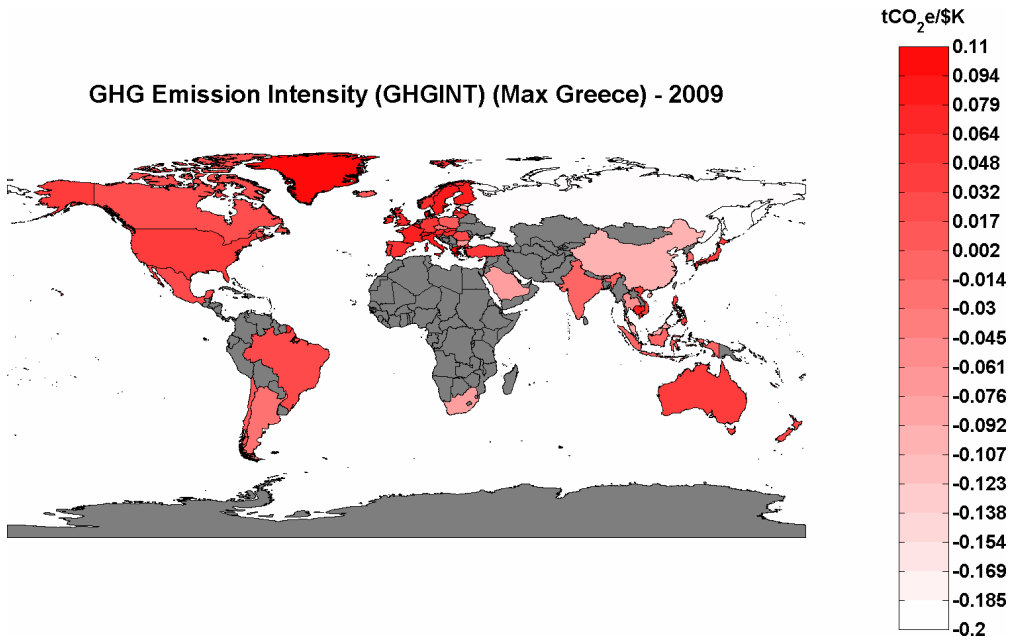

(b) Δ GHGINT

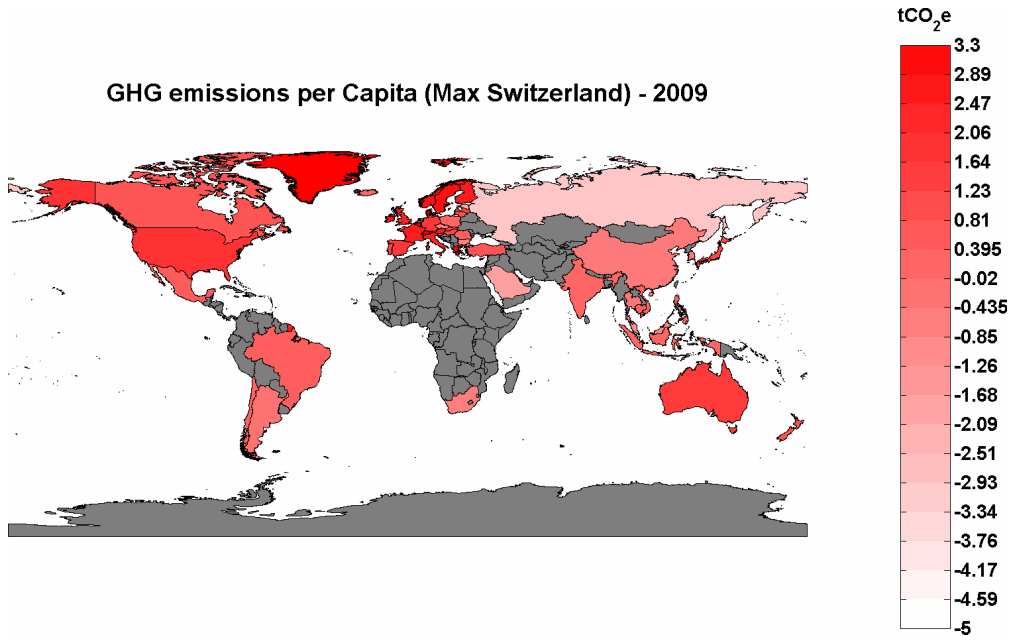

(c) Δ GHGpCapita

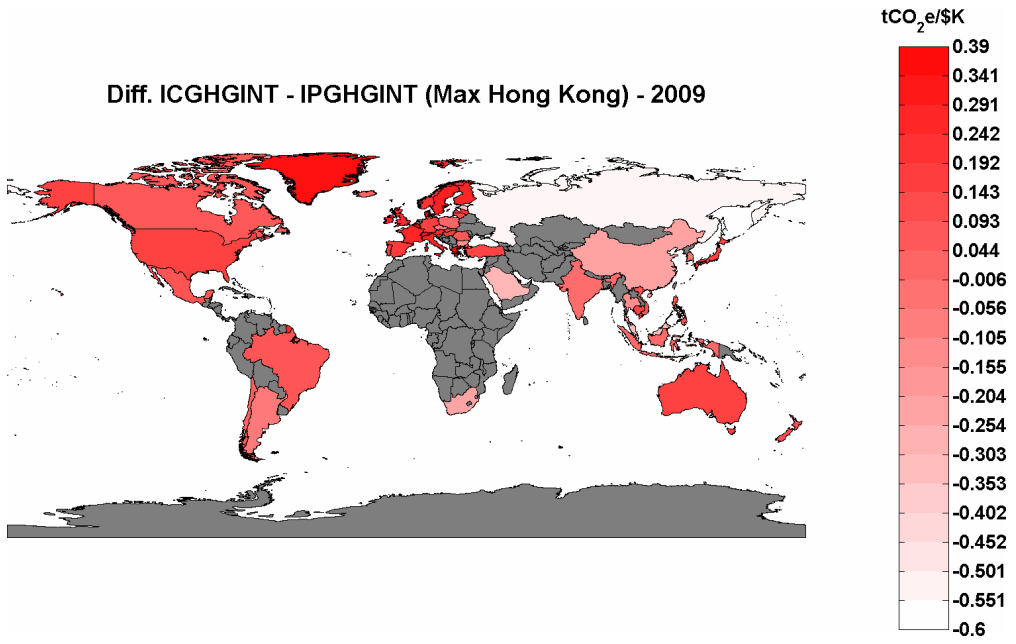

(d) ICGHGINT - IPGHGINT

Figure 4. The visual comparison of the difference in the four indicators for the two cases of consumption-based and production-based $CO_2$ emissions.

| COI | IPGHGINT | COI | IPGHGINT |
|---|---|---|---|
| **Australia** | **2.36** | **Hong Kong** | **0.64** |
| **Brazil** | **1.854** | **India** | **1.265** |
| **Canada** | **1.922** | **Japan** | **1.104** |
| **China** | **2.274** | **Russia** | **2.902** |
| **France** | **0.799** | **South Africa** | **2.847** |
| **Germany** | **1.086** | **Switzerland** | **0.59** |
| **Greece** | **1.052** | **United States** | **1.655** |

(a) IPGHGINT

| COI | ICGHGINT | COI | ICGHGINT |
|---|---|---|---|
| **Australia** | **2.487** | **Hong Kong** | **1.033** |
| **Brazil** | **1.91** | **India** | **1.234** |
| **Canada** | **1.971** | **Japan** | **1.223** |
| **China** | **2.034** | **Russia** | **2.333** |
| **France** | **1.038** | **South Africa** | **2.592** |
| **Germany** | **1.226** | **Switzerland** | **0.876** |
| **Greece** | **1.406** | **United States** | **1.785** |

(b) ICGHGINT

Table 2. The IPGHGINT and the ICGHGINT of the COIs

As mentioned in [3], the GHG emissions intensity indicators cannot be directly used to determine possible carbon border tax or adjustment (BCT/BCA) rates to drive the regions economy toward a low emission operation. In order to estimate the admissible emissions of a region, and in turn its emission credits or debit, we follow the same approach in [3] by considering a green and a red scenario for the world emissions (please see section 2. for more details). In the green scenario, the world total admissible emissions, $\text{GB1}_{2009}$, are 35.55 $\text{GtCO}_2\text{e}$ for 2009. The limit of the red scenario, $\text{RA1B}_{2009}$, is 39.17 $\text{GtCO}_2\text{e}$ in the same year. The admissible emissions of a region is calculated by sharing the green scenario's limit among all the regions based on their GHG emissions intensity:

$$\text{ADMEM}_{i,y} = \left(\frac{\text{IHDIGDP}_{i,y}}{\text{IHDIGDP}_y}\right)\text{GB1}_y, \tag{11}$$

where $\text{ADMEM}_{i,y}$ is the admissible emissions of country $i$ in year $y$.

Then, the emission credits and emission debt of a region are calculated as negative and positive difference between their actual emissions and their admissible emissions:

$$\text{EMCRD}_{i,y} = \lfloor \text{ADMEM}_{i,y} - \text{EM}_{i,y}^{\text{CONS}} \rfloor, \tag{12}$$

$$\text{EMDBT}_{i,y} = \lfloor \text{EM}_{i,y}^{\text{CONS}} - \text{ADMEM}_{i,y} \rfloor, \tag{13}$$

where $\text{EMCRD}_{i,y}$ and $\text{EMDBT}_{i,y}$ are the emission credits and emission debt of country $i$ in year $y$, respectively. The $\lfloor\cdot\rfloor$ denotes that the enclosed quantity is equal to itself when its value is positive, and zero otherwise. As can be seen, in contrast to the IPGHGINT, we use the consumption-based emissions in the calculations of the EMCRD and EMDBT.

Finally, the RED percentage of a region is calculated by dividing their emission debt to their emission debt margin, $\text{EMDBT}_{i,y}^{\text{MARG}}$:

$$\text{EMDBT}_{i,y}^{\text{MARG}} = \left(\frac{\text{ADMEM}_{i,y}}{\text{GB1}_y}\right)(\text{RA1B}_y - \text{GB1}_y), \tag{14}$$

$$\text{RED}_{i,y} = 100\left(\frac{\text{EMDBT}_{i,y}}{\text{EMDBT}_{i,y}^{\text{MARG}}}\right), \tag{15}$$

where $\text{RED}_{i,y}$ is the RED percentage of country $i$ in year $y$. The RED percentage of the COIs is provided in Table 3.

The RED percentage (shown as RED% in the tables) is a measure of low efficient polluting of a region, and therefore can be used to calculate that region's associated BCT rate:

$$\text{BCT}_{i,y} = \text{RED}_{i,y}\big/\text{RED}_{\text{BCT}} \tag{16}$$

| Country | RED % | Country | RED % |
|---|---|---|---|
| **Australia** | **584** | **Hong Kong** | **0** |
| **Brazil** | **248** | **India** | **0** |
| **Canada** | **293** | **Japan** | **0** |
| **China** | **526** | **Russia** | **943** |
| **France** | **0** | **South Africa** | **907** |
| **Germany** | **0** | **Switzerland** | **0** |
| **Greece** | **0** | **United States** | **116** |

(a) Production-based

| Country | RED % | Country | RED % |
|---|---|---|---|
| **Australia** | **668** | **Hong Kong** | **0** |
| **Brazil** | **285** | **India** | **0** |
| **Canada** | **325** | **Japan** | **0** |
| **China** | **367** | **Russia** | **566** |
| **France** | **0** | **South Africa** | **738** |
| **Germany** | **0** | **Switzerland** | **0** |
| **Greece** | **0** | **United States** | **202** |

(b) Consumption-based

Table 3. The RED percentage of the COIs for the two cases of production-based (IPGHGINT) and consumption-based (ICGHGINT) $CO_2$ emissions.

where $RED_{BCT}$ is a conversion factor from RED percentage to the BCT rates. This factor is considered in order to avoid global economic instability because of negative growth rates. Following [3], we consider $RED_{BCT} = 100.0$ for 2009 in this work. In this way, the $0.01^{th}$ of the values shown in Table 3 are the actual BCT rates can be imposed on each of the COIs.

Interestingly, as seen from Table 3, there is no country in the COIs that has an on/off status with respect to the BCT rate. However, if we look at all countries in the database, there are six cases where the BCT rate is either switched to zero or became non-zero after considering the consumption-based emissions. In particular, three European countries, Finland, Ireland, and Cyprus, switch to a non-zero BCT of %0.80, %0.50, and %1.50, respectively. At the same time, one European country, Czech Republic, and two Asian countries, Indonesia and Thailand, upgrade to a zero BCT from their initial production-based BCT rates of %0.60, %0.20, and %0.50, respectively. In terms of the COIs, China, Russia, and South Africa are those that observe a (considerable) decrease in their associated BCT rates. This factor shows the great potential of the proposed universal indicators to play as a converged too in the negotiations toward a global agreement.

| COI | EMCRD (MtCO2e) | COI | EMCRD (MtCO2e) |
|---|---|---|---|
| **Australia** | **0** | **Hong Kong** | **0.064** |
| **Brazil** | **0** | **India** | **0.407** |
| **Canada** | **0** | **Japan** | **0.421** |
| **China** | **0** | **Russia** | **0** |
| **France** | **0.423** | **South Africa** | **0** |
| **Germany** | **0.334** | **Switzerland** | **0.079** |
| **Greece** | **0.043** | **United States** | **0** |

(a) Production-based

| COI | EMCRD (MtCO2e) | COI | EMCRD (MtCO2e) |
|---|---|---|---|
| **Australia** | **0** | **Hong Kong** | **0.034** |
| **Brazil** | **0** | **India** | **0.465** |
| **Canada** | **0** | **Japan** | **0.288** |
| **China** | **0** | **Russia** | **0** |
| **France** | **0.275** | **South Africa** | **0** |
| **Germany** | **0.216** | **Switzerland** | **0.054** |
| **Greece** | **0.007** | **United States** | **0** |

(b) Consumption-based

Table 4. The emission credits of the COI for the two cases of production-based and consumption-based $CO_2$ emissions.

For example, France and Germany almost lost %35 of their emission credits when we switch from the IPGHGINT to the ICGHGINT. This shows that even big EU actors could easily slip from their green status and zero-BCT in the future if they do not consider improvements in terms of consumption-driven imported GHG emissions and also their overall production value.

# 5. Inequality-adjusted Production/Consumption-Insensitive GHG Intensity (IIGHGINT) indicator

Although the ICGHGINT indicator seems to be fair by considering the impact of the final consumers, there could be possible cases where a region is highly biased in terms of production or consumption, and as a consequence the ICGHGINT allows that region to receive a low GHG intensity score without improving its status. For example, a region with low a quality of life (consumption) can have a low ICGHGINT value even if its production-based emissions are high. Although these cases could be a few, a new universal indicator is introduced in this section that is more direct in terms of both production and consumption. We can this indicator the Inequality-adjusted Production/Consumption-Insensitive GHG Intensity (IIGHGINT). The IIGHGINT is defined as follows to disaggregate $\text{GDP}^{\text{BAL}}$ and $\text{IHDIxCapita}^{\text{BAL}}$:

$$\text{IIGHGINT}_{i,y} = \frac{Z}{2}\left(\frac{\text{EM}_{i,y}}{\text{GDP}_{i,y}^{\text{BAL}}} + \frac{\text{EM}_{i,y}^{\text{CONS}}}{\text{IHDIxCapita}_{i,y}^{\text{BAL}}}\right), \tag{17}$$

where $Z$ is the same normalization factor defined in Equation (5). The first term in the definition of the IIGHGINT measure the efficiency of production-based emissions to the production output ($\text{GDP}^{\text{BAL}}$) of the country, while the second term evaluate its performance with respect to the consumption-based emissions and its "internal" activities ($\text{IHDIxCapita}^{\text{BAL}}$). This disaggregation ensures that the allowance associated to the internal activities are not used for production purposes and vice versa.

The rest of formulation to calculate the BCT rates is also modified as follows:

$$\text{EMCRD}_{i,y} = (1/2)\left\lfloor 2\text{ADMEM}_{i,y} - \text{EM}_{i,y} - \text{EM}_{i,y}^{\text{CONS}} \right\rfloor, \tag{18}$$

$$\text{EMDBT}_{i,y} = (1/2)\left\lfloor \text{EM}_{i,y} + \text{EM}_{i,y}^{\text{CONS}} - 2\text{ADMEM}_{i,y} \right\rfloor, \tag{19}$$

In other words, the average of the production- and consumption-based emissions is used in calculations of the emission credits and debt. The formulations related to the $\text{ADMEM}_{i,y}$ , $\text{EMDBT}^{\text{MARG}}$, and $\text{RED}_{i,y}$ are not changed, and the same equations of (11), (14), (15) in the previous section.

With this mentality, the other indicators are also redefined in order to make them more consistent with their purposes:

$$\text{GHGINT}_{i,y} = \text{PGHGINT}_{i,y} = \frac{\text{EM}_{i,y}}{\text{GDP}_{i,y}^{\text{PPP}}}, \tag{20}$$

$$\text{GHGpCapita}_{i,y} = \frac{\text{EM}_{i,y}^{\text{CONS}}}{\text{Capita}_{i,1990}}, \tag{21}$$

Although we do not suggest using these indicators, it is suggested to use the aforementioned definition in case these indicators are used. The GHGINT can be indirectly used as a measure of production efficiency, and the GHGpCapita can be used as an indicator of development efficiency. However, the dynamic nature of societies across the world and high level of in-place connectivity (and to be added) makes it impossible to assume a fixed consumption behavior in various countries, and a balanced status quo between GHGINT and GHGpCapita performances of a country can rapidly breaks down. Therefore, it is suggested that the IIGHGINT indicator is used, which not only gives a combined picture of production and consumption of a region but also can be used to infer on potential changes in its GHG intensity performance.

A comparison among the IPGHGINT, the ICGHGINT, and the IIGHGINT of the COIs is presented in Table 5. Hong Kong in the COI has the highest rate of increase in the GHG intensity of %122 when switching from the ICGHGINT to the IIGHGINT. Australia replaces South Africa as the worst performance in terms of the IIGHGINT with 3.65 $tCO_2e/\$K$.

As can be seen from the table, China has a seems-unrecoverable relative difference of %500 compared to the best European country in the COIs, Switzerland. In contrast, the IIGHGINT indicator was able to provide a more realistic and fair picture, and the associated relative difference between China and Switzerland reduces to a more reasonable value of %78. This shows the great potential of the IIGHGINT indicator as an universal indicator of GHG performance without giving unlimited allowance to developing countries – that could lead to hard to detect emission leakages – while accounting for the consumption-based responsibilities of all counties.

| COI | GHGINT* | COI | GHGINT* |
|---|---|---|---|
| Australia | 1.041 | Hong Kong | 0.253 |
| Brazil | 1.015 | India | 1.024 |
| Canada | 0.849 | Japan | 0.469 |
| China | 1.576 | Russia | 1.699 |
| France | 0.37 | South Africa | 1.448 |
| Germany | 0.511 | Switzerland | 0.26 |
| Greece | 0.499 | United States | 0.7 |

(a) Normalized GHGINT

| COI | IPGHGINT | COI | IPGHGINT |
|---|---|---|---|
| Australia | 2.36 | Hong Kong | 0.64 |
| Brazil | 1.854 | India | 1.265 |
| Canada | 1.922 | Japan | 1.104 |
| China | 2.274 | Russia | 2.902 |
| France | 0.799 | South Africa | 2.847 |
| Germany | 1.086 | Switzerland | 0.59 |
| Greece | 1.052 | United States | 1.655 |

(b) IPGHGINT

| COI | ICGHGINT | COI | ICGHGINT |
|---|---|---|---|
| Australia | 2.487 | Hong Kong | 1.033 |
| Brazil | 1.91 | India | 1.234 |
| Canada | 1.971 | Japan | 1.223 |
| China | 2.034 | Russia | 2.333 |
| France | 1.038 | South Africa | 2.592 |
| Germany | 1.226 | Switzerland | 0.876 |
| Greece | 1.406 | United States | 1.785 |

(c) ICGHGINT

| COI | IIGHGINT | COI | IIGHGINT |
|---|---|---|---|
| Australia | 3.65 | Hong Kong | 2.289 |
| Brazil | 2.033 | India | 1.278 |
| Canada | 2.898 | Japan | 1.992 |
| China | 2.154 | Russia | 2.654 |
| France | 1.3 | South Africa | 3.077 |
| Germany | 1.533 | Switzerland | 1.21 |
| Greece | 1.661 | United States | 2.968 |

(d) IIGHGINT

Table 5. The comparison of four GHG emissions intensities for the COIs.

After repeating all the calculation related to the emission allowance, emission credits, and emission debit, the RED% and BCT rates can be caluclated in the case of the IIGHGINT indicator. Table 6 provides a comparison of the BCT rates considering all three universal indicators. As expected, the IIGHGINT case provides reasonable rates that could motivate the associated countries to place proper actions toward increasing production performance, reducing GHG emissions, reducing unsustainable consumption, and increasing development and quality of life. In particular, comparing the IIGHGINT to the ICGHGINT, South Africa, Russian Federation, and China receive increase in the BCT rates. In contrast, comparing the IIGHGINT to the IPGHGINT, Australia, Canada, Brazil, and the USA see an increase. As mentioned before, many of those COIs that hold zero-BCT rates have received considerable decrease in their emission credits when the ICGHGINT or the IIGHGINT indicators are used, and therefore they are practically in the marginal area and could easily move to non-zero-BCT group in the future. This shows the great potential to improve the GHG emission related status of all countries across the world toward ubiquitous development, efficiency, and sustainability.

| COI | tax % (a) IPGHGINT-based | tax % (b) ICGHGINT-based | tax % (c) IIGHGINT-based |
|---|---|---|---|
| **Australia** | **5.8** | **6.7** | **6.3** |
| **Brazil** | **2.5** | **2.9** | **2.7** |
| **Canada** | **2.9** | **3.3** | **3.1** |
| **China** | **5.3** | **3.7** | **4.5** |
| **France** | **0** | **0** | **0** |
| **Germany** | **0** | **0** | **0** |
| **Greece** | **0** | **0** | **0** |
| **Hong Kong** | **0** | **0** | **0** |
| **India** | **0** | **0** | **0** |
| **Japan** | **0** | **0** | **0** |
| **Russia** | **9.4** | **5.7** | **7.5** |
| **South Africa** | **9.1** | **7.4** | **8.2** |
| **Switzerland** | **0** | **0** | **0** |
| **United States** | **1.2** | **2** | **1.6** |

(a) IPGHGINT-based (b) ICGHGINT-based (c) IIGHGINT-based

Table 6. The BCT rates for the COIs considering three universal GHG emissions intensity indicators.

It is worth noting that, in the definition of the IIGHGINT, we could use a geometrical mean.[9] That definition would be more aggressive toward the lower performance with respect to the production and consumption. In contrast, we chose not to use the geometrical mean in order to give a more positive picture to every region and allow them to leverage on their stronger performance and improve the weak one.

Finally, we want to emphasize on the importance of an EEBT analysis. As an example, Australia as the world's leading coal exporter, has exported 136.4 Mt thermal coal and also 125.0 Mt metallurgical in 2009.[10] Considering the GHG emission factors of $CO_2$, $CH_4$, and $N_2O$ of 2832.3, 0.8, and 5.4, and 2381.4, 0.8, 5.4 in $KtCO_2e/Mt$ for thermal and metallurgical coals, respectively,[11] the Australia's coal exports carried a potential of $680.48(=354.81+325.67)$ $MtCO_2$ GHG emissions in 2009, which is equivalent to %120 of their production-based GHG emissions in the same year.[12] Assuming that %14.5 of these exports was directed to China[13], the share in the production-based GHG emissions of China would be 98.67 $MtCO_2e$, i.e., %1 of the China's emissions.[14] The modification of the IIGHGINT indicator in order to include the EEBT analysis will be consider in the

[9] The geometrical-mean IIGHGINT could be defined as:

$$\text{IIGHGINT}^{g}_{i,y} = 2Z\left(\frac{\text{EM}_{i,y}}{\text{GDP}^{\text{BAL}}_{i,y}}\right)\left(\frac{\text{EM}^{\text{CONS}}_{i,y}}{\text{IHDIxCapita}^{\text{BAL}}_{i,y}}\right)\Bigg/\left(\frac{\text{EM}_{i,y}}{\text{GDP}^{\text{BAL}}_{i,y}}+\frac{\text{EM}^{\text{CONS}}_{i,y}}{\text{IHDIxCapita}^{\text{BAL}}_{i,y}}\right),$$

.

[10] http://www.ret.gov.au/resources/mining/australian_mineral_commodities/coal/Pages/australia_coal_industry.aspx

[11] http://www.climatechange.gov.au/sites/climatechange/files/documents/07_2013/national-greenhouse-accounts-factors-july-2013.pdf

[12] This potential GHG emissions content was equivalent to %114 of their consumption-based emissions.

[13] http://www.australiancoal.com.au/facts-and-figures.html

[14] For Japan, the main importer of the Australia's coal with a share of %39.3,[13] the content can be estimated to 267.42 $MtCO_2e$, equivalent to a considerable amount (%21.6) of Japan's emissions in 2009.

future studies.

## 6. Conclusions

In continuation of the road map toward universal indicators of GHG emissions intensity, the inequality-adjusted consumption-based GHG intensity (ICGHGINT) indicator has been introduced. The ICGHGINT indicator includes the impact of consumption related emissions, and therefore is more resilience to possible carbon and emissions leakage problems. The impact of this indicator has been analyzed using the OECD's STAN IO GHG database, and it has been observed that many regions with green performance against the production-based indicator, the IPGHGINT, fall short against the ICGHGINT. In particular, the EU countries seem to be vulnerable to consumption-related emissions while their industrial sectors suffer from tight regulations. In another step forward, and in order to prevent possible misplacement of production and consumption emissions, the inequality-adjusted production/consumption-insensitive GHG intensity (IIGHGINT) indicator has been introduced. The disaggregative nature of this indicator provides a consistent improvement in both production efficiency and human development in all regions, from the consume-oriented developed countries to production-oriented developing countries. The performance and analytic advantages of the IIGHGINT are compared with the other indicators, including the GHGINT, the IPGHGINT, and the ICGHGINT, and it has been shown that this universal indicator can provide a fair and resolvable picture of all regions.

As future prospective, inclusion of the geographical and climate advantages of regions in the calculations of their GHG emissions intensity will be considered. This will provide a means to avoid these advantages mask the low performance of a region in terms of efficiency and development. In addition, improvement of the GHG emissions intensity indicator by considering the emissions embodied in bilateral trade (EEBT) will be considered. Finally, in order to estimate the observable impact of every of these universal indicators on the global market and trades, and in turn on the emissions of every regions across the world, a Multi-Region Input-Output (MRIO) analysis will be performed considering the proposed BCT rates compared to the business as usual of absence of such BCT rates.

## Acknowledgments

The authors thank the NSERC of Canada for their financial support under grant CRDPJ 424371-11.